\documentclass{elsart}
\usepackage{graphicx,amssymb}
\usepackage{epsfig}
\usepackage{graphicx}

\journal{Physica A}

\begin{document}

\begin{frontmatter}
\title{Relaxation in homogeneous and non-homogeneous polarized systems. A mesoscopic entropy approach}
\author{J. G. M\'endez-Berm\'udez$^{\dagger}$, I. Santamar\'ia-Holek$^{\ddagger}$}
\address{$^{\dagger}$Instituto de F\'isica "Luis Rivera Terrazas", BUAP. \\ Av. San Claudio y 18 Sur, San Manuel. 72570, Puebla,
M\'{e}xico.\\}
\address{$^{\ddagger}$Facultad de Ciencias, Universidad Nacional Aut\'{o}noma de M\'{e}xico.\\
Circuito exterior de Ciudad Universitaria. 04510, D. F.,
M\'{e}xico.}
\thanks[fn1]{E-mail: isholek.fc@gmail.com}

\begin{abstract}
The dynamics of a degree of freedom associated to an axial vector
in contact with a heat bath is described by means of a probability
distribution function obeying a Fokker-Planck equation. The equation is derived by
using mesoscopic non-equilibrium thermodynamics and permits a formulation of a
dynamical theory for the axial degree of freedom
(orientation, polarization) and its associated order parameter. The theory is used to describe
dielectric relaxation in homogeneous and non-homogeneous systems in the
presence of strong electric fields. In the homogeneous case, we obtain the dependence
of the relaxation time on the external field as observed in experiments. In the
non-homogeneous case, our model accounts for the two observed maxima of the dielectric loss
giving a good quantitative description of experimental data at all frequencies, especially for
systems with low molecular mass.
\end{abstract}
\end{frontmatter}
\section{Introduction}

Orientational dynamics of a nematic director or of a dipole in an applied electric field
plays a crucial role in numerous modern technologies. For example,
the reorientation of the director
vector in nematic liquid-crystals changes the electro-optical properties of these materials, making
of central interest the response of these systems to constant and
time varying electrical fields \cite{gennes}.
Liquid crystals, polymer solutions and dielectrics are systems showing
strong anisotropies in their physical properties such as viscosities, translational and rotational diffusion
coefficients, and dielectric or magnetic susceptibilities
\cite{gennes,doi,onsager,hijar,MNETpolim,mendoza1,dejardin0,coffey,dejardin,malaga}.
This fact together with the interaction of these systems with their environment, leads to a number of phenomena
not observed in isotropic phases, like the dependence of the characteristic relaxation times
of the reorientation of molecules on the intensity of an externally applied electrical field \cite{yin,block-hayes}.

From a fundamental viewpoint, the interaction with the surroundings
and the orientational relaxation of the system can be modelled in terms of the
dynamics of an axial degree of freedom $\hat{n}$ in contact with a heat bath \cite{gennes,doi}.
Due to the importance of thermal fluctuations in the dynamics of the axial vector, in this article we
begin our analysis by calculating the mesoscopic entropy production $\sigma$ of the system
using the rules of mesoscopic nonequilibrium
thermodynamics \cite{r-nmet}. Then, from the entropy production and the probability conservation, we derive a
Fokker-Planck equation for the probability density $f(\vec{r},\hat{n},t)$ which depends on
time $t$, position $\vec{r}$, and the axial vector $\hat{n}$. This Fokker-Planck equation
allows us to formulate evolution equations for the average of the axial vector and for the associated order
parameter in two important physical situations: \textit{i}) When the systems can be considered
homogeneous and \textit{ii}) when non-homogeneities are important. The results we have
obtained can be compared with experiments \cite{block-hayes,adachi0} and previous theories based
on other approaches \cite{block-hayes,ullman,langevinp}.

The paper is organized as follows. In Sec. 2 we calculate the mesoscopic
entropy production, derive the Fokker-Planck
equation for the probability density and obtain the
evolution equations for the axial degree of freedom and its associated order parameter
under the assumption of the system's homogeneity. In Sec. 3 we apply our results to the case
of dielectric relaxation and in Sec. 4 we compare with experiments and with other theories.
In sec. 5 we analyze the relaxation dynamics in non-homogeneous systems by taking into account the
diffusion of the polarization. Finally, in the last section we summarize our main results.

\section{Mesoscopic entropy and Fokker-Planck dynamics}

Consider a system of particles in contact with a heat bath at constant
temperature $T$ and in the presence of an external potential $U$. The state
of the system is characterized by the position vector $\vec{r}$ and the axial
degree of freedom $\hat{n}$ which may be associated to an
orientation or a polarization vector.

The mesoscopic dynamics of the system
can be described in terms of the single-particle probability density
$f(\vec{r},\hat{n},t)$ satisfying the normalization condition $\int f
d\hat{n}d\vec{r}=1$. This condition imposes that $f$ must obey
the probability conservation equation \cite{kroger}
\begin{equation}
\frac{\partial f}{\partial t}= -\frac{\partial}{\partial \vec{r}}
\cdot \left(f \vec{V} \right) -\vec{\mathcal{R}} \cdot \left( f
\vec{\Omega}  \right), \label{cont}
\end{equation}
where $\vec{\mathcal{R}}=\hat{n} \times
\frac{\partial}{\partial \hat{n}}$ is the rotational operator and
$\vec{V}$ is the velocity conjugated to the position vector $\vec{r}$.
Eq. (\ref{cont}) can be obtained by taking into account the corresponding boundary
and periodic conditions over $\vec{r}$ and $\hat{n}$ after applying the triple product identity and
considering that the change in time of the axial vector $\dot{\hat{n}}$ is perpendicular
to $\hat{n}$: $\dot{\hat{n}}=\vec{\Omega} \times \hat{n}$, with $\vec{\Omega}$
the angular velocity \cite{doi}.

The explicit expressions for the unknown probability currents $ f\vec{V} $
and $f\vec{\Omega}$ in $(\vec{r},\hat{n})$-space, can be found by calculating the entropy
production $\sigma$ of the system during its evolution in time. To
this end, we will follow the rules of mesoscopic nonequilibrium thermodynamics which
makes use of the Gibbs entropy postulate (GEP) \cite{r-nmet,mazur}
\begin{equation}
\rho s(t)=-k_{B}\int f\ln\left( \frac{f}{f_{eq}}\right) d\hat{n}+\rho s_{eq},\label{gep}
\end{equation}
where $k_{B}$ is the Boltzmann constant, $\rho=m\int f
d\hat{n}$ is the mass density with $m$ the mass of a particle and
$s_{eq}$ is the  equilibrium entropy. $f_{eq}$ is the probability density
characterizing the equilibrium state of the system, given by
\cite{zwan,debye,hansen}
\begin{equation}
f_{eq}\left( \vec{r},\hat{n}  \right)=f_{0}e^{-m U/k_{B}T},
\label{fle}
\end{equation}
where, $U=U\left( \vec{r},\hat{n} \right)$ is a potential energy
field per unit of mass and $f_{0}$ is a normalization constant.

The GEP formulated in (\ref{gep}) is compatible
with the Gibbs equation \cite{mazur}
\begin{equation}
T \delta s = \delta e + p\delta\left(\rho^{-1}\right) - m\int
\tilde{\mu} \delta \left( \frac{f}{\rho} \right) d\hat{n}d\vec{r},
\label{ge}
\end{equation}
where $s$, $e$, $p$ and $\tilde{\mu}(\vec{r},\hat{n},t)$ are the
non-equilibrium entropy, energy, pressure and chemical potential
per unit mass of the system. Taking the variation of
Eq. (\ref{gep})  and comparing with (\ref{ge}) keeping constant $e$ and $\rho$,
we may identify the nonequilibrium chemical potential
\begin{equation}
\tilde{\mu} = \tilde{\mu}_{eq}+\frac{k_{B}T}{m} \ln \frac{f}{f_{0}} + U, \label{potquim}
\end{equation}
where $\tilde{\mu}_{eq}$ is the chemical potential in the equilibrium state.
The balance equation for the non-equilibrium entropy follows from the
Gibbs equation (\ref{ge}) (with $e$ and $\rho$ constant) by taking the time derivative,
integrating by parts and using the corresponding boundary and periodic conditions for the currents.
We obtain
\begin{equation}
\rho \sigma=-\frac{m}{T}\int\left[ \frac{\partial
\tilde{\mu}}{\partial\vec{r}} \cdot \left( f \vec{V} \right)+
\vec{\mathcal{R}}\tilde{\mu} \cdot \left( f \vec{\Omega}  \right)
\right]d\hat{n}d\vec{r}. \label{produc-s}
\end{equation}
Eq. (\ref{produc-s}) contains two contributions of the type flux-force pair,
each one coming from the diffusion process in $\vec{r}$-
and $\hat{n}$-space, respectively. Following nonequilibrium
thermodynamics rules, linear phenomenological
relationships between currents and thermodynamic forces can be established in the form
\cite{r-nmet,mazur}
\begin{eqnarray}
\vec{V} & = & - \mathbb{L}_{rr} \cdot \frac{\partial\tilde{\mu}}{\partial\vec{r}} -
\mathbb{L}_{rn} \cdot \vec{\mathcal{R}}\tilde{\mu},\label{flujo1}\\
\vec{\Omega} & = & - \mathbb{L}_{nn} \cdot
\vec{\mathcal{R}}\tilde{\mu} - \mathbb{L}_{nr} \cdot
\frac{\partial \tilde{\mu}}{\partial\vec{r}}.\label{flujo2}
\end{eqnarray}
Here, $\mathbb{L}_{rr}$, $\mathbb{L}_{rn}$, $\mathbb{L}_{nr}$ and
$\mathbb{L}_{nn}$  are the (tensorial) Onsager coefficients
which are, in general, functions of $\vec{r}$,
$\hat{n}$, $T$ and $\rho$, \cite{r-nmet,mazur}. However, for simplicity we will
assume here that the Onsager tensors have constant components. Since
$\hat{n}$ is an axial vector and $\vec{r}$ a polar vector, the
Onsager-Casimir relationships impose the relation $\mathbb{L}_{nr}
= - \mathbb{L}_{rn}$, \cite{mazur,gya}.

Substituting Eqs. (\ref{flujo1}) and (\ref{flujo2}) into (\ref{cont}),
we finally arrive at the Fokker-Planck equation
\begin{eqnarray}
\frac{\partial f}{\partial t} &=& \frac{\partial}{\partial\vec{r}}
\cdot \left[ \frac{k_{B}T}{m} \mathbb{L}_{rr} \cdot \left(
\frac{\partial f}{\partial\vec{r}} + \frac{m f}{k_{B}T}
\frac{\partial U}{\partial\vec{r}} \right) \right]\nonumber\\
&& +
\vec{\mathcal{R}} \cdot \left[ \frac{k_{B}T}{m} \mathbb{L}_{nn}
\cdot \left( \vec{\mathcal{R}} f + \frac{m f}{k_{B}T}
\vec{\mathcal{R}} U \right)\right]. \label{f-p1}
\end{eqnarray}
Eq. (\ref{f-p1}) constitutes the basis of the mesoscopic description of the system
and enables one to derive the evolution equation of the moments
of the probability density. A similar equation has been obtained by using
hydrodynamic methods in Ref. \cite{doi} and from a Langevin approach in the case of
a mean field approximation describing inhomogeneous nematics \cite{ilg}. The
present derivation allows the generalization of Eq. (\ref{f-p1}) to cases in which inertia effects
could be important, \cite{agustin}.

\subsection{Relaxation equations for the average value of the axial degree of freedom and the order parameter}

In order to derive the relaxation equations for the
average axial vector $\vec{N}$ and the order parameter $\mathbb{S}$, we will assume
that the system is homogeneous, that is, the
interaction potential of the particles with an external force only
depends on $\hat{n}$, $\tilde{U}=U(\hat{n})$. In this case, the translational
part of Eq. (\ref{f-p1}) can be integrated over $\hat{n}$, yielding a free diffusion equation for
the center of mass of the particles from which the evolution
equation for the second moment of position, $\langle\vec{r}^{2}
\rangle=\int\vec{r}^{2}f(\vec{r},\vec{n},t)d\vec{n}d\vec{r}$, is
\begin{equation}\label{10}
\frac{\partial \left\langle r^{2} \right\rangle }{\partial t} =
2\mbox{Tr}[\mathbb{D}_{tr}].
\end{equation}
Here, $\mbox{Tr}[\mathbb{D}_{tr}]$ is the trace of the
translational diffusion tensor $\mathbb{D}_{tr}$ assumed constant for simplicity
and which is related to the Onsager coefficient $\mathbb{L}_{rr}$  through the
expression $\mathbb{D}_{tr}=(k_{B}T/m)\mathbb{L}_{rr}$, \cite{doi}. Eq. (\ref{10}) implies that the center of
mass of the particles performs Brownian motion with second moment
$\left\langle r^{2} \right\rangle=2\mbox{Tr}[\mathbb{D}_{tr}]t$.
In the case of solid systems at sufficiently low temperatures, the
translational diffusion coefficient can be taken as zero.


Thus, by integrating Eq. (\ref{f-p1}) over $\vec{r}$ we obtain the
following Fokker-Planck equation for the reduced probability distribution
$g\left( \hat{n},t \right)=\int fd\vec{r}$
\begin{equation}
\frac{\partial g}{\partial t} = \vec{\mathcal{R}} \cdot \left[ \frac{k_{B}T}{m} \mathbb{L}_{nn} \cdot \left( \vec{\mathcal{R}} g + \frac{m g}{k_{B}T} \vec{\mathcal{R}}\tilde{U} \right) \right].
\label{dg/dt}
\end{equation}
In the present case, the Onsager coefficient $\mathbb{L}_{nn}$ is
related with the matrix of rotational relaxation times
$\underline{\gamma}$ through $\mathbb{L}_{nn}=
\frac{m}{k_{B}T}\underline{\gamma}^{-1}$, which defines
the rotational diffusion coefficient. From
Eq. (\ref{dg/dt}), we can obtain the relaxation equations for the
first and second-order moments of $g\left( \hat{n},t \right)$ corresponding to
the average axial vector $\vec{N}$ and the order parameter $\mathbb{S}$, defined by \cite{doi}
\begin{equation}
\vec{N}(t)= \int \hat{n}gd\hat{n} \,\,\,\,\,\,\,\, and\,\,\,\,\,\,\,\, \mathbb{S}(t)=\langle {\bf S} \rangle(t) = (1/2)\int (3\hat{n}\hat{n}-{\bf 1})gd\hat{n},
\label{nprom}
\end{equation}
where ${\bf 1}$ is the unit tensor and  ${\bf S}=(1/2)(3\hat{n}\hat{n}-{\bf 1})$ is the microscopic order parameter.
It is convenient to mention that when $\hat{n}$ represents the orientation vector of a molecule, then
$\mathbb{S}$ is the orientation tensor \cite{doi,gennes} that is
related to the well known Onsager order parameter characterizing
the order-disorder transition in suspensions of rod-like molecules \cite{onsager}.
For polar systems, $\hat{n}$ and $\mathbb{S}$ are
related to the polar and quadrupolar terms of the corresponding multipolar expansion.

Now, taking the time derivative of equation (\ref{nprom}) and
using (\ref{dg/dt}), after integrating by parts assuming periodic conditions for the
probability current, one obtains the
evolution equation
\begin{equation}
\frac{\partial \vec{N}}{\partial t} = -\underline{\tilde{\gamma}}^{-1}\cdot
\left[
\vec{N}
-\frac{2m}{3k_{B}T}\underline{\tilde{\gamma}}\cdot\left\langle\mathbb{M}({\bf S})\cdot\frac{\partial U}{\partial\hat{n}}\right\rangle
+\frac{m}{3k_{B}T}\left\langle\frac{\partial U}{\partial\hat{n}}\right\rangle
\right],
\label{dn/dt2}
\end{equation}
where $\mathbb{M}({\bf S})$ is defined by
\begin{equation}
\mathbb{M}({\bf S})=
\left[\underline{\tilde{\gamma}}^{-1}\cdot{\bf S}
+\left( \underline{\gamma}^{-1}:{\bf S}\right){\bf 1}
-\underline{\gamma}^{-1}\cdot {\bf S}\right].
\label{m1}
\end{equation}
In Eq. (\ref{dn/dt2}), the matrix of relaxation times is given by $\underline{\tilde{\gamma}}^{-1}=\mbox{Tr}(\underline{\gamma}^{-1}){\bf 1}-\underline{\gamma}^{-1}$, implying that, in the general case,
the characteristic relaxation times are combinations of the elements of the rotational diffusion
tensor. The first term on the right-hand side of Eq. (\ref{dn/dt2}) comes from
the diffusion term of Eq. (\ref{dg/dt}), whereas the other
terms are due to the external force. Eq. (\ref{dn/dt2})
is expected to be valid for characteristic values of the interaction energy, $U$, of the order
of the thermal energy $k_BT$.

Following a similar procedure,
Eqs. (\ref{dg/dt}) and (\ref{nprom}) yield the evolution
equation for $\mathbb{S}$
\begin{eqnarray}
\frac{\partial\mathbb{S}}{\partial t}&=&
-2\mathbb{M}^{s}({\mathbb{S}})-2\left(\underline{\tilde{\gamma}}^{-1}\cdot\mathbb{S}\right)^{s}
+\frac{6m}{5k_{B}T}\left\langle\tilde{\mathbb{M}}({\bf Q})\cdot\frac{\partial U}{\partial\hat{n}}\right\rangle^{s}\nonumber\\
&&-\frac{3m}{5k_{B}T}\left[2\left\langle\left(\underline{\tilde{\gamma}}^{-1}\cdot\frac{\partial U}{\partial\hat{n}}\right)\hat{n}\right\rangle^{s}
+\left\langle\underline{\tilde{\gamma}}^{-1}\left(\hat{n}\cdot\frac{\partial U}{\partial\hat{n}}\right)\right\rangle^{s}\right]\nonumber\\
&&+\frac{3m}{5k_{B}T}\left[\left\langle\left(\underline{\tilde{\gamma}}^{-1}\cdot\frac{\partial U}{\partial\hat{n}}\right)\cdot\hat{n}\right\rangle {\bf 1}
+\left\langle \left(\underline{\gamma}^{-1}\cdot\hat{n}\right)\frac{\partial U}{\partial\hat{n}}\right\rangle^{s}\right],
\label{ds/dt2}
\end{eqnarray}
where the superscript $s$ stands for the symmetric part of a tensor and
$\mathbb{M}(\mathbb{S})$ has the same form as in Eq. (\ref{m1}) but in (\ref{ds/dt2}) depends
on the averaged order parameter $\mathbb{S}$. In Eq. (\ref{ds/dt2}) we have introduced the
average tensor $\mathbb{Q}$ defined by
\begin{equation}
\mathbb{Q}(t)=\langle{\bf Q}\rangle(t)=(1/2)\int\left[5\hat{n}\hat{n}\hat{n}-\left(\hat{n}{\bf 1}\right)^{s}\right]gd\hat{n},
\label{moment3}
\end{equation}
and the tensor
\begin{equation}
\tilde{\mathbb{M}}({\bf Q})=
\left[\underline{\tilde{\gamma}}^{-1}\cdot{\bf Q}
+\left(\underline{\tilde{\gamma}}^{-1}:{\bf Q}\right){\bf 1}
-\underline{\gamma}^{-1}\cdot{\bf Q}\right].
\label{m2}
\end{equation}
Here,  ${\bf Q}$ is the microscopic octupolar tensor and $\left(\hat{n}{\bf 1}\right)^{s}_{ijk}=n_{i}\delta_{jk}+n_{j}\delta_{ki}+n_{k}\delta_{ij}$.
In Eq. (\ref{ds/dt2}), the first three terms come from the diffusion term in Eq. (\ref{dg/dt})
whereas the others are due to interactions with the external field.
Eqs. (\ref{dn/dt2}) and (\ref{ds/dt2}) constitute the first
two evolution equations of the complete hierarchy of moments
of the distribution function $g$.

\section{Dipole and quadrupole relaxation in a homogeneous system}


The relaxation of the polarization vector in dielectrics can be described
by means of equations (\ref{dn/dt2}) and (\ref{ds/dt2}) since
$\hat{n}$ can be interpreted as the dipole moment of a molecule
through the expression: $\hat{n}=\mu^{-1}\vec{p}$ where $\mu$ is the magnitude of the
dipole moment of the molecule. In this form, the polarization $\vec{P}$ and
the cuadrupolar moment $\mathbb{S}_{p}$ per unit volume are defined by
\cite{jack}
\begin{equation}
\vec{P}=N\mu \vec{N}\hspace{0.5cm}\mbox{and}\hspace{0.5cm}\mathbb{S}_{p}=N^{2}\mu^{2}\left\langle {\bf S}\right\rangle \label{polar1}
\end{equation}
where the tensor $\mathbb{S}_{p}$ is the quadrupole moment induced by thermal
fluctuations. In first approximation, the
potential energy per unit mass of a polar molecule interacting
with an external electric field $\vec{E}$ is given by
\cite{doi}
\begin{equation}
\tilde{U}(\vec{p})=-\frac{\vec{p}\cdot\vec{E}}{m}.
\label{u1}
\end{equation}

From Eqs. (\ref{dn/dt2}), (\ref{ds/dt2}) and (\ref{polar1}), it follows that the
evolution equation for $\vec{P}$ is
\begin{eqnarray}
\frac{\partial\vec{P}}{\partial t}&=& -\underline{\tilde{\gamma}}^{-1}\cdot
\left[\vec{P}+\frac{2}{3Nk_{B}T}\underline{\tilde{\gamma}}\cdot\mathbb{M}(\mathbb{S}_{p})\cdot\vec{E}
-\frac{N\mu^{2}}{3k_{B}T}\vec{E}\right],
\label{dn/dte1}
\end{eqnarray}
where $\mathbb{M}(\mathbb{S}_{p})$ is defined according to Eq. (\ref{m1}). The evolution equation for $\mathbb{S}_{p}$ takes the form
\begin{eqnarray}
\frac{\partial\mathbb{S}_{p}}{\partial t}&=&
-2\mathbb{M}^{s}(\mathbb{S}_{p})-2\left(\underline{\tilde{\gamma}}^{-1}\cdot\mathbb{S}_{p}\right)^{s}
-\frac{6}{5Nk_{B}T}\left[\tilde{\mathbb{M}}(\mathbb{Q}_{p})\cdot\vec{E}\right]^{s}\nonumber\\
&&+\frac{3N\mu^{2}}{5k_{B}T}\left\{2\left[\left(\underline{\tilde{\gamma}}^{-1}\cdot\vec{E}\right)\vec{P}\right]^{s}
+\left[\underline{\tilde{\gamma}}^{-1}\left(\vec{P}\cdot\vec{E}\right)\right]^{s}\right\}\nonumber\\
&&-\frac{3N\mu^{2}}{5k_{B}T}\left\{\left[\left(\underline{\tilde{\gamma}}^{-1}\cdot\vec{E}\right)\cdot\vec{P}\right]{\bf 1}
+\left[\left(\underline{\gamma}^{-1}\cdot\vec{P}\right)\vec{E}\right]^{s}\right\},
\label{dn/dts1}
\end{eqnarray}
where $\mathbb{Q}_{p}=N^{3}\mu^{3}\mathbb{Q}$ with $\tilde{\mathbb{M}}(\mathbb{Q}_{p})$ defined according
to Eq. (\ref{m2}). The contribution of higher order moments will be neglected for
convenience. These equations generalize the Debye
theory of dielectric relaxation by taking into
account the coupling between the dipolar and quadrupolar moments
naturally contained in the distribution function $g$. Debye's theory is recovered
if quadrupolar contributions are neglected in Eqs. (\ref{dn/dte1}) and (\ref{dn/dts1}).

A closed set of equations for $\vec{P}$ and $\mathbb{S}_{p}$ can
be obtained by approximating $\mathbb{Q}_{p}$ by its equilibrium average,
which can be calculated using the equilibrium distribution function
\begin{equation}
g_{eq}=Z^{-1}e^{m\tilde{U}/k_{B}T}. \label{geqloc4}
\end{equation}
Eq. (\ref{geqloc4}) is the stationary solution of (\ref{dg/dt}) and
$Z^{-1}=\int e^{m\tilde{U}/k_{B}T}d\vec{p}$ is the partition function of the system.
Notice that, when the externally applied field is weak, it is usual to take
the so-called homogeneous isotropic approximation of (\ref{geqloc4}),
given by \cite{doi}
\begin{equation}
g_{eq}\cong\frac{1}{4\pi}.
\label{geqloc3}
\end{equation}
We will use Eqs. (\ref{geqloc4}) and (\ref{geqloc3}) in order to complete the set of equations
(\ref{dn/dte1}) and (\ref{dn/dts1}) and to make
comparisons with previous theories.

\subsection{Dielectric relaxation in the homogeneous-isotropic approximation}

By using Eqs. (\ref{dn/dte1}) and
(\ref{dn/dts1}), and the
isotropic distribution function (\ref{geqloc3}), the dielectric response of a polar system
will be analyzed in terms of the complex dielectric susceptibility $\chi$.

To proceed, we will now assume that the external electric field is time
dependent and given by the sum
of a strong constant electric field $\vec{E}_{0}$ applied along the $z$ axis, and a weak
AC probe field $\vec{E}_{1}(t)=\vec{E}_{1}e^{i\omega_{\mathrm{o}} t}$ applied
in an arbitrary direction, that is: $\vec{E}=\vec{E}_{0}+\vec{E}_{1}(t)$.
In addition, we will assume for
simplicity that $\partial \mathbb{S}_{p}/\partial t\sim 0$ and
solve Eqs. (\ref{dn/dte1})  and (\ref{dn/dts1}) for $\vec{P}$. This quasi-stationary
assumption leads to a considerable simplification of the obtained formulas
and do not modifies considerably the results of our
model.  Thus, since the average of the octupolar tensor (\ref{moment3}) with (\ref{geqloc3})
vanishes ($[\mathbb{Q}_{p}]_{ijk}=0$), from Eq. (\ref{dn/dts1}) we obtain the
following relation for $\mathbb{S}_{p}$
\begin{eqnarray}
\mathbb{M}^{s}(\mathbb{S}_{p})+2\left(\underline{\tilde{\gamma}}^{-1}\cdot\mathbb{S}_{p}\right)^{s}&=&
\frac{3N\mu^{2}}{5k_{B}T}\left\{2\left[\left(\underline{\tilde{\gamma}}^{-1}\cdot\vec{E}(t)\right)\vec{P}\right]^{s}
+\left[\underline{\tilde{\gamma}}^{-1}\left(\vec{P}\cdot\vec{E}(t)\right)\right]^{s}\right.\nonumber\\
&&-\left.\left[\left(\underline{\gamma}^{-1}\cdot\vec{P}\right)\vec{E}(t)\right]^{s}
-\left[\left(\underline{\tilde{\gamma}}^{-1}\cdot\vec{E}(t)\right)\cdot\vec{P}\right]{\bf 1}\right\}.
\label{ds/dtc11}
\end{eqnarray}
From Eqs. (\ref{ds/dtc11}) and (\ref{dn/dte1}), it is possible to obtain two equations for
the parallel $P_{\|}(t)$ and perpendicular $P_{\bot}(t)$ components of the polarization vector defined with respect
to the constant field $\vec{E}_{0}$. Hence, using the condition $E_1 << E_0$ in the second term of Eq. (\ref{dn/dte1}) and keeping dominant terms, the equation for $P_{\|}(t)$ becomes
\begin{equation}
\frac{\partial P_{\|}(t)}{\partial t}=-\tau^{-1}_{\|}\left[P_{\|}(t)-\epsilon_\mathrm{o}\chi_{\|}E_{\|}(t)\right], \label{dppar/dt}
\end{equation}
where $E_{\|}=E_0+E^{\|}_{1}$ with $E^{\|}_{1}$ the component of $\vec{E}_1$ paralel to $\vec{E}_0$,
and we have assumed that the matrix of relaxation
times $\underline{\gamma}$ is diagonal with components $\gamma_{\bot}$, $\gamma_{\bot}$
and $\gamma_{\|}$. By definition, $\underline{\tilde{\gamma}}$ and $\underline{\gamma}$ have the same symmetries.
The relaxation time $\tau_{\|}$ and the zero frequency
susceptibility $\chi_{\|}$ in Eq. (\ref{dppar/dt}) are given by
\begin{equation}
\tau_{\|}=\frac{\gamma_{\bot}/2}{1+\frac{\xi^{2}}{30}\left(4-\Gamma\right)} \hspace{1cm}\mbox{and}\hspace{1cm} \chi_{\|}=\frac{\chi_{\mathrm{o}}}{1+\frac{\xi^{2}}{30}\left(4-\Gamma\right)}.
\label{chitaupar1}
\end{equation}
Here, we have introduced $\Gamma\equiv \gamma_{\bot}/\gamma_{\|}$ to simplify the notation
and defined $\chi_\mathrm{o}\equiv N\mu^{2}/3\epsilon_\mathrm{o}k_{B}T$ with $\epsilon_{\mathrm{o}}=8.854\times10^{-12}$F/m, \cite{jack}.
It is important to notice the fact that the constant electric field introduces corrections to
$\tau_{\|}$ and $\chi_{\|}$ through the dimensionless parameter $\xi=\mu E_0/k_{B}T$.

Using Fourier transform techniques, from Eq. (\ref{dppar/dt}) one obtains
\begin{equation}
P_{\|}(t)=\epsilon_\mathrm{o}\chi_{\|}E_\mathrm{o} + \epsilon_\mathrm{o}\chi_{\|}(\omega_{\mathrm{o}})E^{\|}_{1}(t),
\label{p-par-1}
\end{equation}
where $\chi_{\|}(\omega_{\mathrm{o}})=\chi_{\|}(1+i\omega_\mathrm{o}\tau_{\|})^{-1}$.
The evolution equation for the perpendicular component of the polarization vector can be calculated in similar
form. One obtains
\begin{equation}
\frac{\partial P_{\bot}(t)}{\partial
t}=-\gamma^{-1}_{\bot}\left[P_{\bot}(t)-\epsilon_\mathrm{o}\chi_\mathrm{o}E_{1}^{(\bot)}(t)\right],
\label{dpper/dt}
\end{equation}
where $E_{1}^{(\bot)}$ is the component of $\vec{E}_1$
perpendicular to $\vec{E}_0$. Following the same procedure to obtain
(\ref{chitaupar1}), we find that:
$P_{\bot}=\epsilon_{\mathrm{o}}\chi_{\bot}(1+i\omega_{\mathrm{o}}\tau_{\bot})^{-1}E_{1}^{(\bot)}(t)$.
In this case, $\tau_{\bot}$ and $\chi_{\bot}$ take the form
\begin{equation}
\tau_{\bot}=\frac{\gamma_{\bot}/(1+\Gamma)}{1+\frac{\xi^{2}}{10}\frac{(2-\Gamma)}{(1+\Gamma)}}
\hspace{0.5cm}\mbox{and}\hspace{0.5cm}
\chi_{\bot}=\frac{\chi_{o}}{1+\frac{\xi^{2}}{10}\frac{(2-\Gamma)}{(1+\Gamma)}}.
\label{chitauper1}
\end{equation}
These results yield the following expressions for the real and imaginary parts
of the parallel and perpendicular components of the complex dielectric susceptibility tensor
\begin{eqnarray}
&\chi'_{\|}(\omega_\mathrm{o})=
\frac{\chi_{\|}}{1+\omega_\mathrm{o}^{2}\tau_{\|}^{2}}+\chi^\infty_{\|}
\hspace{1cm} \mbox{and} \hspace{1cm}
\chi''_{\|}(\omega_\mathrm{o})=\frac{\omega_{\mathrm{o}}\tau_{\|}\chi_{\|}}{1+\omega_{\mathrm{o}}^{2}\tau_{\|}^{2}}
,\label{chipar2}&\\
&\chi'_{\bot}(\omega_\mathrm{o})=\frac{\chi_{\bot}}{1+\omega_{\mathrm{o}}^{2}\tau_{\bot}^{2}}+ \chi^\infty_{\bot} \hspace{1cm} \mbox{and}\hspace{1cm} \chi''_{\bot}(\omega_{\mathrm{o}})=\frac{\omega_{\mathrm{o}}\tau_{\bot}\chi_{\bot}}{1+\omega_{\mathrm{o}}^{2}
\tau_{\bot}^{2}}.\label{chiper2}&
\end{eqnarray}
where $\chi_{\|}^{\infty}$ and $\chi_{\bot}^{\infty}$ are the high frequency susceptibilities of
the system. From Eqs. (\ref{chitaupar1}) and (\ref{chitauper1}), it follows that the quadrupole moment
induced by thermal flucutations introduces field dependent
corrections in the susceptibility and the relaxation times, as observed in experiments \cite{block-hayes}.

We have also calculated the expressions for the parallel and perpendicular susceptibilities of the
system in the non-stationary case. The expressions for the components of the polarization vector
can only be given in Fourier space and, in the isotropic approximation they are given by
\begin{equation}\label{ne-1}
\hat{P}_{\|}^{ns}(\omega)=\epsilon_{o}\frac{\chi_{\|}^{ns}}{1-i\omega c_{\|}\tau_{\|}^{ns}}\hat{E}_{\|}(\omega)
\hspace{0.5cm} \mbox{and}\hspace{0.5cm}
\hat{P}_{\bot}^{ns}(\omega)=\epsilon_{o}\frac{\chi_{\bot}^{ns}}{1-i\omega c_{\bot}\tau_{\bot}^{ns}}\hat{E}_{\bot}(\omega),
\end{equation}
where the superscript $ns$ indicates the non-stationary case. The relaxation times $\tau_{\|}^{ns}$ and $\tau_{\bot}^{ns}$ are now given by
\begin{equation}
\tau_{\|}^{ns}=\frac{\gamma_{\bot}/2}{1+\frac{\xi^{2}}{30}\frac{\left(4-\Gamma\right)}{\left[1+\left(\omega\tau_{\bot}/6\right)^{2}\right]}}
\hspace{0.5cm}\mbox{and}\hspace{0.5cm}
\tau_{\bot}^{ns}=\frac{\gamma_{\bot}/(1+\Gamma)}{1+\frac{\xi^{2}}{10}\frac{\left(2-\Gamma\right)}{(1+\Gamma)\left[1+\left(\omega\tau_{\bot}/6\right)^{2}\right]}},
\label{ne-3}
\end{equation}
whereas the susceptibilities $\chi_{\|}^{ns}$ and $\chi_{\bot}^{ns}$  in this non-stationary case are defined by
\begin{equation}
\chi_{\|}^{ns}=\frac{\chi_{o}}{1+\frac{\xi^{2}}{30}\frac{\left(4-\Gamma\right)}{\left[1+\left(\omega\tau_{\bot}/6\right)^{2}\right]}}
\hspace{0.5cm}\mbox{and}\hspace{0.5cm}
\chi_{\bot}^{ns}=\frac{\chi_{o}}{1+\frac{\xi^{2}}{10}\frac{\left(2-\Gamma\right)}{(1+\Gamma)\left[1+\left(\omega\tau_{\bot}/6\right)^{2}\right]}}.
\label{ne-4}
\end{equation}
Finally, the correction factors $c_{\|}$ and $c_{\bot}$ due to the time derivative in Eq. (\ref{dn/dts1})
are respectively given by
\begin{eqnarray}
c_{\|}&=&1-\frac{\xi^{2}}{90}\frac{\left(4-\Gamma\right)}{\left[1+\left(\omega\tau_{\bot}/6\right)^{2}\right]},
\nonumber\\
c_{\bot}&=&1-\frac{\xi^{2}}{60}\frac{\left(2-\Gamma\right)}{\left[1+\left(\omega\tau_{\bot}/6\right)^{2}\right]},
\label{ne-2}
\end{eqnarray}
In the following section, the comparison with experiments using the complete set of non-stationary relations
for the susceptibilities,  Eqs. (\ref{ne-1})-(\ref{ne-2}), or their non-homogeneous versions (see Sec. 5),
will show that the quantitative description is very similar to the one given by means of the stationary expressions,
that are simpler.

\subsection{Relaxation in an homogeneous non-isotropic approximation}

Eqs. (\ref{chitaupar1}) and (\ref{chitauper1}) for the relaxation times and the susceptibilities
have the correct qualitative dependence on the external field $E_0$  as observed in experiments,
but they fail quantitatively. This is due to the fact that the isotropic approximation is not adequate
when strong electrical fields are applied to the system.

Therefore, the quantitative description can be improved by taking the average (\ref{moment3})
in Eqs. (\ref{dn/dte1}) and (\ref{dn/dts1}) with the non-isotropic probability distribution (\ref{geqloc4}).
In this case, the components of the tensor $\mathbb{Q}_{p}$ do not vanish and more complicated formulas
are obtained for $\mathbb{S}_{p}$ and $\vec{P}$. Thus, following a procedure similar to that of subsection 3.1,
it is possible to derive the following expressions for the zero frequency
parallel and perpendicular susceptibilities
\begin{eqnarray}
\tilde{\chi}_{\|}&=&\frac{\chi_\mathrm{o}\left\{1+\frac{2\xi}{5}(3-2\Gamma)\left[\left(1+\frac{15}{\xi^{2}}\right)L[\xi]-\frac{5}{\xi}\right]\right\}} {1+\frac{\xi^{2}}{30}\left(4-\Gamma\right)},\nonumber\\
\tilde{\chi}_{\bot}&=&\frac{\chi_\mathrm{o}\left[1+\frac{\xi}{5}\left(\frac{2-\Gamma}{1+\Gamma}\right)\left\{L[\xi]-2\Gamma\left[\left(1+\frac{15}{\xi^{2}}\right)L[\xi]
-\frac{5}{\xi}\right]\right\}\right]}{\left[1+\frac{\xi^{2}}{10}\frac{(2-\Gamma)}{(1+\Gamma)}\right]}.
\label{factors4}
\end{eqnarray}
Here, the tilde stands for the non-isotropic case. The expression for the relaxation times do not change
and are given by the corresponding expressions in (\ref{chitaupar1}) and (\ref{chitauper1}).
If we approximate our description by decoupling  Eqs. (\ref{dn/dte1}) and (\ref{dn/dts1}) by
taking the average of $\mathbb{S}_{p}$ with (\ref{geqloc4}), then the expressions for the components of
the susceptibility reduce to
\begin{equation}
\tilde{\chi}_{\|}^{bh}=\frac{3\chi_{\mathrm{o}}L[\xi]}{\xi}\hspace{1cm}  \mbox{and} \hspace{1cm}
\tilde{\chi}_{\bot}^{bh}=\chi_\mathrm{o}\left[1+\left(\frac{2-\Gamma}{1+\Gamma}\right)\left(1-\frac{3L[\xi]}{\xi}\right)\right],
\label{factors3}
\end{equation}
where the parallel component $\tilde{\chi}_{\|}^{bh}$ is equal to the one reported in Ref. \cite{block-hayes}. In this approximation, the parallel and perpendicular relaxation times become: $\tau_{\|}=\gamma_{\bot}/2$ and $\tau_{\bot}=\gamma_{\bot}/[1+\Gamma]$.

In this section we have shown that the field dependent corrections of the
relaxation times and susceptibilities of a dielectric are due to
the coupling between the evolution equations for the dipolar, quadrupolar and,
in general, higher order
moments of the distribution. The results obtained
agree with previous theories (see, for instance, \cite{langevinp}) and, when
$\xi \sim 0$, reduce to the ones derived by means of Debye's theory.

When considering the non-stationary dynamics of the quadrupole, we obtain that in the present case
the susceptibilities entering in the corresponding Eqs. (\ref{ne-1}) are now given by
\begin{eqnarray}
\tilde{\chi}_{\|}^{ns}&=&\chi_{o}\frac{\left\{1+\frac{\xi}{5}\frac{(3-2\Gamma)}{(1-i\omega\tau_{\bot}/6)}\left[\left(1+\frac{15}{\xi^{2}}\right)L[\xi]-\frac{5}{\xi}\right]\right\}}
{1+\frac{\xi^{2}}{30}\frac{\left(4-\Gamma\right)}{\left[1+\left(\omega\tau_{\bot}/6\right)^{2}\right]}},\nonumber\\
\tilde{\chi}_{\bot}^{ns}&=&\chi_{o}\frac{\left\{1+\frac{\xi}{5}\frac{(2-\Gamma)}{(1+\Gamma)(1-i\omega\tau_{\bot}/6)}\left\{L[\xi]-\Gamma\left[\left(1+\frac{15}{\xi^{2}}\right)L[\xi]-\frac{5}{\xi}\right]\right\}\right\}}
{1+\frac{\xi^{2}}{10}\frac{\left[\left(2-\Gamma\right)/\left(1+\Gamma\right)\right]}{\left[1+\left(\omega\tau_{\bot}/6\right)^{2}\right]}}.
\label{ne-5}
\end{eqnarray}
The relaxation times and the correction factors $c_{\|}$, $c_{\bot}$ are still given by Eqs. (\ref{ne-3}) and
(\ref{ne-2}). In the case when there is no coupling we have $c_{\|}=c_{\bot}=1$, and therefore
the relaxation times and the susceptibilities are given reduce to the form Eq. (\ref{factors3}), as expected.
\begin{figure}[]
\centerline{\includegraphics[width=6cm]{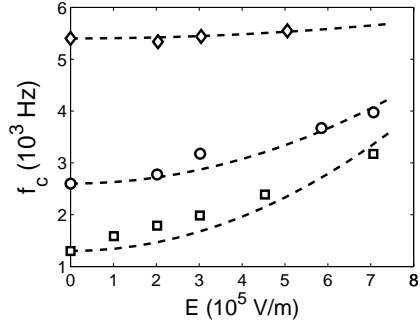}}
\caption{Comparison of the experimental data of the
critical frequency $f_{c}(E)$ (symbols) taken from Ref.
\cite{block-hayes} with the theoretical results for a strong DC field
(lines) for a dilute solution of PBLG molecules with molecular
weights $M_{w}$ of 1.1 $\times 10^{5}$(romboids), 2.6 $\times
10^{5}$(circles) and 4.6 $\times 10^{5}$(squares).}
\label{frecc}
\end{figure}

\section{Comparison with experiments}

In this section, we use  Eq. (\ref{chitaupar1})  for the relaxation time
to estimate the value of the dipolar moment of PBLG molecules with different
molecular masses \cite{block-hayes}. In similar form, the
susceptibilities given in (\ref{factors4}) are used to fit the experimental data
for the dielectric loss of {\it cis}-PI polymers in solution reported in
Ref. \cite{adachi1}.

\subsection{Determination of the dipolar moment using the field dependent relaxation time}

The magnitude of the dipolar moment $\mu$ of a molecule can be estimated by
using the critical frequency $f_{c}$, defined
as the value of the frequency at which the maximum of the dielectric loss occurs.
Since the value of $f_{c}$ depends on the magnitude of the applied field, see Fig. 1,
then $\mu$  can be determined by using the relation \cite{coffey,adachi1}
\begin{equation}
f_{c}=(2\pi\tau_{\|})^{-1}. \label{frecctau}
\end{equation}

The open symbols in Fig. 1 show isothermal (298 K) data of the critical
frequency $f_{c}$, reported from experiments with dilute solutions of PBLG having different
molecular masses $M_{w}$, \cite{block-hayes}. The data were obtained by
applying a strong DC electric field ($0 -9\times 10^{5}$ V/m) on the system and
can be fitted using Eq. (\ref{chitaupar1}) by assuming
that the energy of a dipole in the DC field is of the order of the
thermal energy ($\xi \approx 1$), and that $\gamma_{\bot}\sim\gamma_{\|}=\gamma$.
In this case the relaxation time takes the form
$\tau_{\|}=(\gamma/2)/[1+(\xi^{2}/10)]$.

\begin{table}[]
\begin{center}
\scriptsize{
\begin{tabular}{|c|c|c|c|c|c|c|c|}
\hline
\, $M_{w}(\times 10^{5})$ \,  & \, Debye \,  & \, BH-1 \, & \, BH-2 \, &  \, WKC \,  & \, Ullman \, & FP \\
\hline
\, 4.6 \, & \, 8.7 \, & \, 7.1 \, & \, 26 \, & \, 23.3 \, & \, 22.3 \, & \, 23.2 \, \\
\hline
\, 2.6 \, & \, 5.6 \, & \, 4.6 \, & \, -  \, & \, 14.7 \, & \,  - \,  & \, 13.9 \, \\
\hline
\, 1.1 \, & \, 2.8 \,  & \, 2.3 \, & \,  - \, & \, 4.8 \, & \, - \, & \, 4.0 \\
\hline
\end{tabular}
}
\end{center}
\vspace{0.3cm} \caption{Values of the dipolar moment
$\mu$ (10$^{-27}$C m) of a dilute solution of PBLG
polar molecules having different molecular masses: $M_{w}= 1.1, 2.6$ and $4.6
\times 10^{5}$. The first line denotes different models used for comparison (see text
for details).  The last column contains the results obtained with our
model. }\label{table1}
\end{table}

Table 1 shows a comparison of the values of $\mu$ obtained by using different theories.
The values estimated with the Debye and the Block and Hayes model 1
(BH-1), were obtained in the weak-field approximation.
For the system with the largest molecular mass in the strong-field approximation, the magnitude
of the dipolar moment obtained with the Block and Hayes model 2 (BH-2) is very similar to those
obtained by means of a theory based on the Langevin equation (WKC) \cite{langevinp}, and a classical result
reported by Ullman in Ref. \cite{ullman}.
The last column in Table 1 shows that the results obtained with our model (FP) in the
non-isotropic (strong field) approximation are in good agreement with the experimental results.

\subsection{Normal and $\alpha$ relaxations in type-C polymers}

\begin{figure}[!]
\par
\centering \mbox{\includegraphics[width=5.5cm]{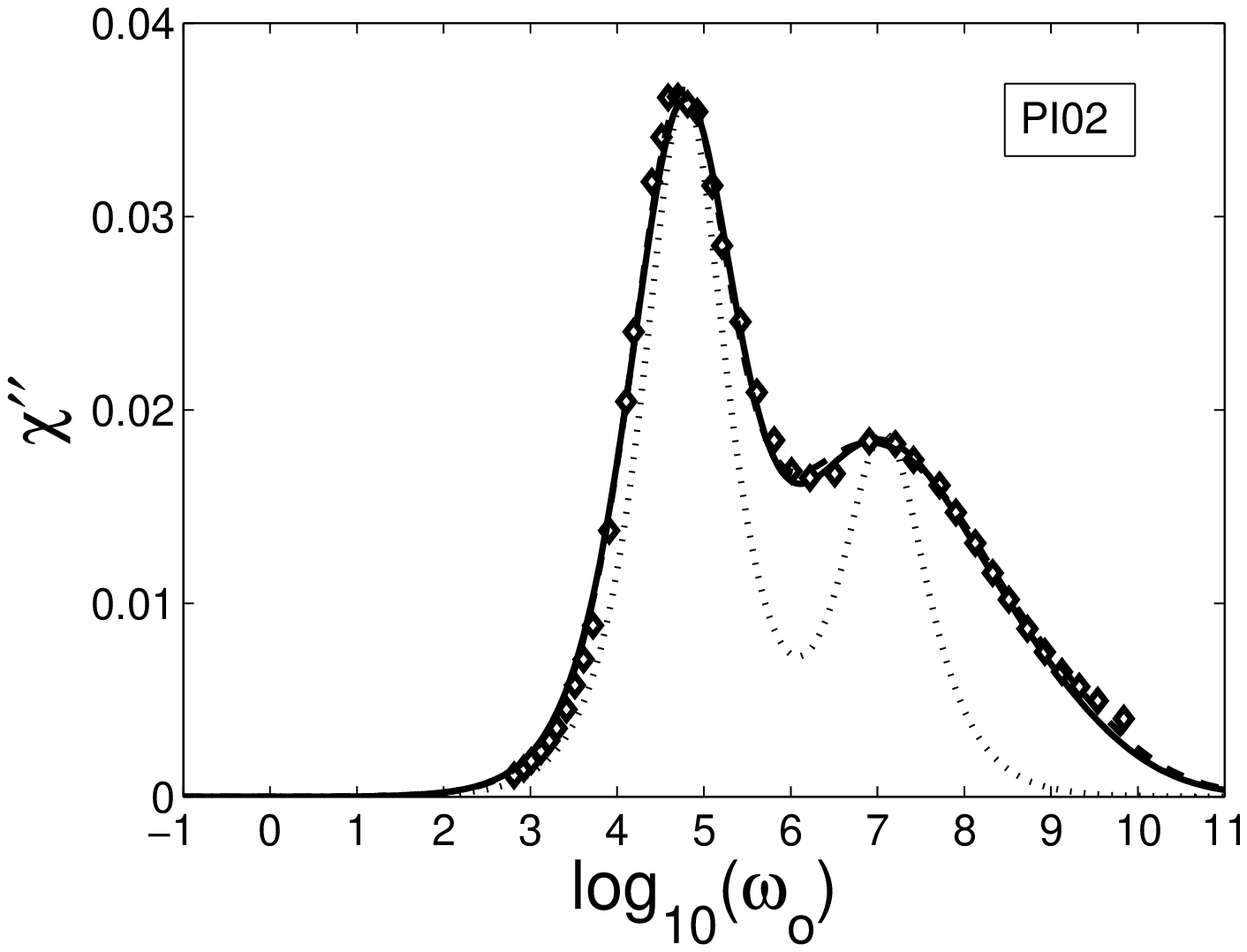}}
\mbox{\includegraphics[width=5.5cm]{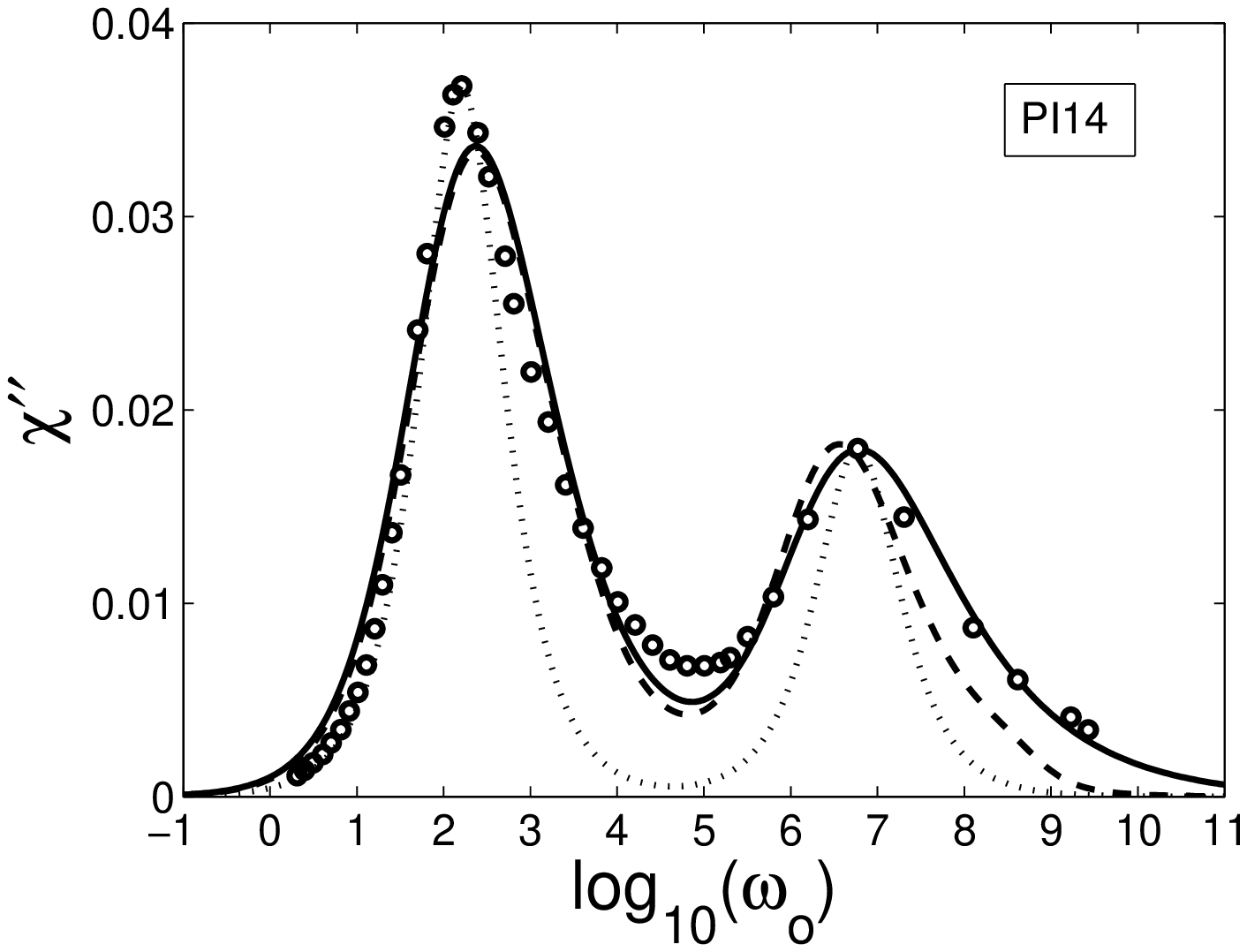}}
\mbox{\includegraphics[width=5.5cm]{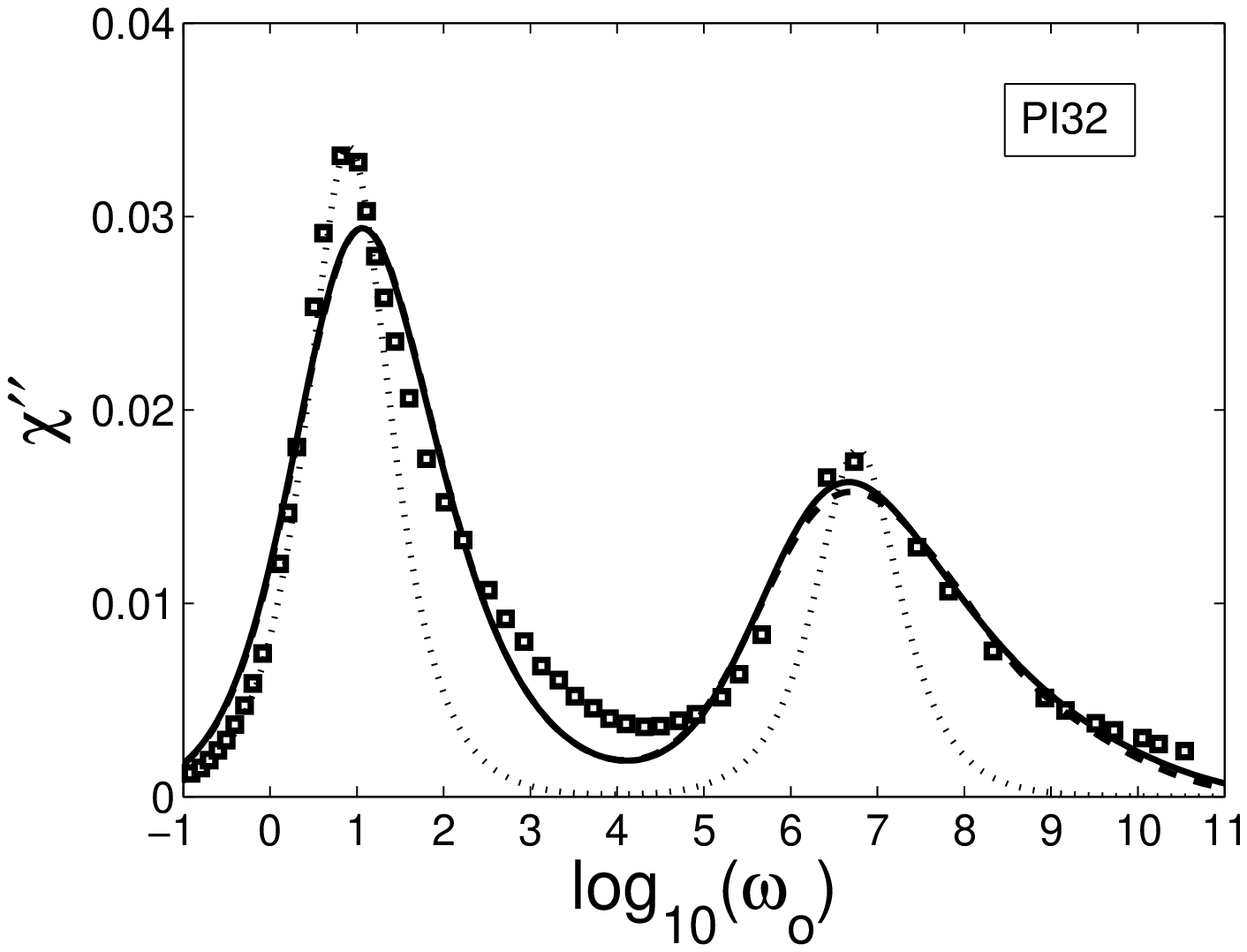}}
\par
\caption{Experimental data (symbols) \cite{adachi1} of the dielectric loss of
highly concentrated solution of {\it cis}-PI chains at 273 K as a function of
frequency for different mass weights $M_{w}$. The average molecular mass in kg-mol$^{-1}$ are:
a) 2; b) 14 and c) 32. The dotted lines correspond to the best fit
obtained with Eqs. (\ref{chipar2}) and (\ref{chiper2}) under the
assumption of an homogeneous system. The solid lines correspond to
the best fit using Eqs. (\ref{eq9}) and (\ref{eq10}) for the
non-homogeneous approximation where we have used dispersion relations
of the form $k^2(\omega)= (\omega_{\mathrm{o}}/\omega^*)^{\alpha}$  for the parallel
and perpendicular components of the translational diffusion coefficient. Here,
$\alpha\neq 1$ and $\omega^*$ is a characteristic frequency associated with the
viscoelastic nature of the medium (see details in Sec. 5). The dashed lines correspond to the
non-stationary and non-homogeneous approximation given by Eqs. (\ref{ne-6}) and (\ref{ne-7}).} \label{loss1}
\end{figure}

Highly concentrated solutions of {\it cis}-polyisoprene  ({\it cis}-PI)
show a dielectric relaxation that involves both normal and $\alpha$-relaxations
characterized by two absorption peaks of the imaginary part of the dielectric susceptibility;
see Fig. \ref{loss1} and Ref. \cite{adachi1}.
The first maximum of the dielectric loss is usually attributed to
the dipolar moment aligned with the chain contour, and is associated to the normal relaxation
characteristic of Type-A polymers. The second
maximum is related to the dipolar moment perpendicular to the chain contour
and characterizes Type-B polymers. The {\it cis}-PI presents both types of
relaxations and therefore is classified as a Type-C polymer.

For this system, the polarization $\vec{p}$ can be interpreted as a local degree of freedom in
contact with a heat bath, and therefore its dielectric relaxation can be described
by using Eqs. (\ref{chipar2})-(\ref{factors4}) only after assuming
that the system is homogeneous (the translational diffusion of
the polarization is neglected).

The comparison between experiments (symbols) and theory (lines) shown
in Fig. \ref{loss1}, was performed by taking the total imaginary part of the
susceptibility as the sum of the parallel $\tilde{\chi}''_{\|}(\omega_{\mathrm{o}})$
and perpendicular $\tilde{\chi}''_{\bot}(\omega_{\mathrm{o}})$ contributions given in (\ref{chipar2})
and (\ref{chiper2}), respectively. The low frequency maximum of $\tilde{\chi}''$ was used
to obtain the values of $\xi$, $\chi_0$ and $\tau_{\|}$ as was done in section 4.1. Then,
after determining the characteristic frequency ($\tau_{\bot}^{-1}$) at
which the high frequency maximum appears, Eqs. (\ref{chipar2})
and (\ref{chiper2}) were used to fit the whole set of experimental data.
From the dotted line in Fig. \ref{loss1}, it is clear that our model gives an amplitude
of the second maximum which coincides well with the
experimental one for the three molecular masses of the \textit{cis}-PI considered.
Table 2 contains the values of the parameters used to fit the data. Notice that,
in this case, the interaction energy per molecule between the system and the
external constant field is about five times larger than the thermal energy.
Since only two relaxation modes were considered, the theoretical curves are
narrower than the experimental ones, as expected. In the following section we will show that,
to obtain broader curves, one has to consider the translational diffusion of polarization.
\begin{table}[]
\begin{center}
\scriptsize{
\begin{tabular}{|c|c|c|c|c|c|c|}
\hline
Polymer & $\chi_{0}$ & $\xi$ & $\gamma_{\|}\,[s]$ & $\gamma_{\bot}\,[s]$  & $\tau_{\|}\,[s]$ & $\tau_{\bot}\,[s]$ \\
\hline
PI02 & 0.13 & 7.48 & 1.66$\times10^{-5}$ & 8.25$\times10^{-8}$ & 1.96$\times10^{-6}$ & 6.77$\times10^{-9}$ \\
\hline
PI14 & 0.13 & 7.98 & 6.18$\times10^{-3}$ & 1.69$\times10^{-7}$ & 6.51$\times10^{-4}$ & 1.23$\times10^{-8}$ \\
\hline
PI32 & 0.12 & 7.07 & 1.24$\times10^{-1}$ & 1.69$\times10^{-7}$ & 1.62$\times10^{-2}$ & 1.54$\times10^{-8}$ \\
\hline
\end{tabular}
}
\end{center}
\vspace{0.3cm}\caption{\label{table2} Values of the parameters
obtained with Eqs. (\ref{chipar2}) and (\ref{chiper2}) in the
homogeneous approximation (see dotted lines in Fig. \ref{loss1}).
$\chi_{0}$ and $\xi$ are dimensionless quantities.}
\end{table}

\section{Diffusion effects in dielectric relaxation}

The previous description of dielectric relaxation in \textit{cis}-PI solutions
can be improved by considering
the effects of non-homogeneities of the dipolar and
the quadrupolar moments. Therefore, it is relevant to consider the local definitions
\begin{equation}
\vec{P}(\vec{r},t)=N\int \vec{p}f(\vec{p},\vec{r},t)d\vec{p};\,\,\,\,\mathbb{S}_p(\vec{r},t)=\frac{N^2}{2}\int
(3\vec{p}\vec{p}-\mu^2{\bf 1})f(\vec{p},\vec{r},t)d\vec{p},\,\,\,\,\,\,
\label{pfield}
\end{equation}
where we have assumed that the distribution function depends on the polarization vector

Similarly as in section 2, the evolution equations for
$\vec{P}(\vec{r},t)$ and $\mathbb{S}_p(\vec{r},t)$ can be calculated by taking
the time derivative of Eqs. (\ref{pfield}),
substituting (\ref{f-p1}) in the resulting expression and integrating by parts
assuming the corresponding boundary conditions. The
resulting diffusion equation for $\vec{P}(\vec{r},t)$ is
\begin{eqnarray}
\frac{\partial\vec{P}}{\partial t} &=&\left(\mathbb{D}_{tr}:\nabla\nabla\right)\vec{P}-\underline{\tilde{\gamma}}^{-1}\cdot\left[\vec{P}+\frac{2}{3Nk_{B}T}\underline{\tilde{\gamma}}\cdot\mathbb{M}(\mathbb{S}_{p})\cdot\vec{E}
-\frac{N\mu^{2}}{3k_{B}T}\vec{E}\right],
\label{eq1}
\end{eqnarray}
whereas for $\mathbb{S}_{p}(\vec{r},t)$ the diffusion equation is
\begin{eqnarray}
\frac{\partial\mathbb{S}_{p}}{\partial t}&=&\left(\mathbb{D}_{tr}:\nabla\nabla\right)\mathbb{S}_{p}-\frac{6}{5Nk_{B}T}\left[\tilde{\mathbb{M}}(\mathbb{Q}_{p})\cdot\vec{E}\right]^{s}
-2\mathbb{M}^{s}(\mathbb{S}_{p})-2\left(\underline{\tilde{\gamma}}^{-1}\cdot\mathbb{S}_{p}\right)^{s}\nonumber\\
&&+\frac{3N\mu^{2}}{5k_{B}T}\left\{2\left[\left(\underline{\tilde{\gamma}}^{-1}\cdot\vec{E}\right)\vec{P}\right]^{s}
+\left[\underline{\tilde{\gamma}}^{-1}\left(\vec{P}\cdot\vec{E}\right)\right]^{s}\right.\nonumber\\
&&\,\,\,\,\,\,\,\,\,\,\,\,\,\,\,\,\,\,\,\,\,\,\,\,\,\,\,\,\,\,\,\,
\left.-\left[\left(\underline{\tilde{\gamma}}^{-1}\cdot\vec{E}\right)\cdot\vec{P}\right]{\bf 1}
-\left[\left(\underline{\gamma}^{-1}\cdot\vec{P}\right)\vec{E}\right]^{s}\right\}.
\label{eq2}
\end{eqnarray}
The first term on the right-hand side of equations (\ref{eq1}) and (\ref{eq2}) accounts for
the translational diffusion, $\nabla\equiv \partial/\partial \vec{r}$ is the gradient operator and we have assumed
that the interaction potential only depends
on the externally applied electric field:
$U=-\vec{p}\cdot\vec{E}_{0}$.  Eqs. (\ref{eq1})
and (\ref{eq2}) are coupled through the term containing the tensor $\mathbb{M}(\mathbb{S}_{p})$.
\begin{table}[]
\begin{center}
\scriptsize{
\begin{tabular}{|c|c|c|c|c|c|c|c|c|c|}
\hline
Polymer & $\chi_{0}$ & $\xi$ & $\gamma_{\|}$(s) & $\gamma_{\bot}$(s) & $\alpha$ & $\beta$  & $\omega_{\|}$(Hz) & $\omega_{\bot}$(Hz) \\
\hline
PI02& 1.48 & 2.51 & 4.10$\times10^{-6}$ & 5.75$\times10^{-8}$ & 1.03 & 0.65 & 2.03$\times10^{-1}$ & 3.39$\times10^{-6}$ \\
\hline
PI14& 8.14 & 2.15 & 2.93$\times10^{-4}$ & 1.00$\times10^{-8}$ & 0.79 & 0.71 & 2.19$\times10^{-3}$ & 3.38$\times10^{-6}$ \\
\hline
PI32& 3.69 & 1.87 & 9.79$\times10^{-2}$ & 3.73$\times10^{-8}$ & 0.80 & 0.65 & 1.10$\times10^{-2}$ & 1.12$\times10^{-6}$ \\
\hline
\end{tabular}
}
\end{center}
\vspace{0.3cm}\label{table3} \caption{The numerical values are
obtained with nonlinear regression with Eqs. (\ref{eq9}) and
(\ref{eq10}) in the non-homogeneous approximation (dashed lines in
Fig. \ref{loss1}). $\chi_{0}$, $\xi$, $\alpha$ and $\beta$ are
dimensionless quantities. The values of the relaxation times $\tau$ can
be obtained by using Eqs. (\ref{chitaupar1}) and (\ref{chitauper1}).}
\end{table}

The expressions for the parallel and perpendicular contributions
of the dielectric susceptibility tensor can be obtained by following a procedure
similar to that of Sec. 3.  That is, by substituting the components of
$\mathbb{S}_{p}$ obtained from Eq. (\ref{eq2}) into (\ref{eq1}),
in the quasi-stationary approximation one finds
\begin{eqnarray}
\chi_{\|}(\vec{k},\omega_{\mathrm{o}})&=&\frac{\tilde{\chi}_{\|}}{1+D_{tr\|}k^{2}\tau_{\|}+i\omega_{\mathrm{o}}\tau_{\|}},
\label{eq9}\\
\chi_{\bot}(\vec{k},\omega_{\mathrm{o}})&=&\frac{\tilde{\chi}_{\bot}}{1+D_{tr\bot}k^{2}\tau_{\bot}+i\omega_{\mathrm{o}}
\tau_{\bot}}. \label{eq10}
\end{eqnarray}
The expressions for the relaxation times $\tau_{\|}$ and $\tau_{\bot}$ are given through
Eqs. (\ref{chitaupar1}) and (\ref{chitauper1}). The relation for the parallel and perpendicular components of the
susceptibility $\tilde{\chi}_{\|}$ and $\tilde{\chi}_{\bot}$ are still given by (\ref{factors4}).

Comparison with experiments can be performed by using
the previous expressions and
assuming that the parallel and perpendicular
components of the diffusion tensor are of the form:
$D_{tr\|}k^{2}=(\omega_{\mathrm{o}}/\omega_{\|})^{\alpha}$ and
$D_{tr\bot}k^{2}=(\omega_{\mathrm{o}}/\omega_{\bot})^{\beta}$. This dependence on frequency
has been suggested in the literature and is related to the viscoelastic
nature of the system  \cite{ivan-hdz}, and discussed in relation to the problem of liquid-crystal like
relaxing amphiphilic systems such as lipid membrane. In these last systems, it was shown that
the non-Debye relaxation manifests a memory feeling submesoscopic dynamics that may arise due to a
reduced system's dimensionality that can be accessed phenomenologically \cite{GadEurophys}.

In the non-stationary case, the expressions obtained for the susceptibilities are
\begin{eqnarray}
\chi_{\|}^{ns}(\vec{k},\omega)&=&\frac{\tilde{\chi}_{\|}^{ns}}{1+D_{tr\|}k^{2}\tau_{\|}^{ns}-i\omega c_{\|}\tau_{\|}^{ns}},
\label{ne-6}\\
\chi_{\bot}^{ns}(\vec{k},\omega)&=&\frac{\tilde{\chi}_{\bot}^{ns}}{1+D_{tr\bot}k^{2}\tau_{\bot}^{ns}-i\omega c_{\bot}\tau_{\bot}^{ns}}, \label{ne-7}
\end{eqnarray}
where all the field dependent factors ($\tau_{\|}^{ns}$, $\tau_{\bot}^{ns}$, $c_{\|}$, $c_{\bot}$,
$\tilde{\chi}_{\|}^{ns}$ and $\tilde{\chi}_{\bot}^{ns}$)
are given by Eqs. (\ref{ne-3}), (\ref{ne-2}) and (\ref{ne-5}).

The solid line in Fig. \ref{loss1} shows the theoretical results in the non-homogeneous
case. The effect of diffusion is to make the peaks broader due to the power law relation
between the wave number and the frquency. They also introduce
two relaxation times associated to $\omega_{\|}^{-1}$ and $\omega_{\bot}^{-1}$.
Fig. 2 shows that our approach is more appropriate for describing systems
made up of particles with low molecular mass, although both
maxima are well described even for systems of particles
having large molecular masses.
Table 3 shows the values of the parameters obtained in the present
approximation after using a nonlinear regression of Eqs.
(\ref{eq9}) and (\ref{eq10}) to the experimental data.
Calculations in progress indicate that the description offered here could be improved by incorporating
the effects of mass dispersion.

The results obtained in this section suggest that the coupled relaxation
equations for the dipolar $\vec{P}$ and quadrupolar $\mathbb{S}_{p}$ fields
constitute a minimal model to describe the dielectric relaxation of
solutions of Type-C polymers.

\section{Conclusions}

We have formulated a mesoscopic model
describing the dynamics of systems characterized by an axial degree of freedom
in contact with a heat bath. The description of the dynamics of this degree of freedom is performed
by means of a Fokker-Planck equation for the nonequilibrium probability
distribution function. The equation, derived by calculating the
mesoscopic entropy production, allows one to formulate a
nonequilibrium description of the system in terms of relaxation
equations for the multipoles, related to the moments of the distribution function.

Applications of the theory to describe dielectric relaxation
in homogeneous and non-homogeneous systems in different approximations allowed us
to compare with previous theories and better elucidate the new contributions of our model.
In the case of dilute polymer suspensions, we found that our theory accounts for the dependence
of the relaxation time on the strong
applied electrical field and gives values for the
dipolar moment of the particles which agree with previous ones.

We used our model to describe the normal and $\alpha$ relaxations
in Type-C polymer solutions. Our results clearly show that, in order to correctly account for the
amplitude of the second maximum of the dielectric loss (associated to the $\alpha$ relaxation),
the minimal model has to take into account the coupled diffusion equations for the
dipolar and quadrupolar fields.
The anisotropic diffusion of these quantities introduces new characteristic
times leading to a better description of the experiments.  In particular, we found that
our theory is very useful in describing dielectric relaxation of solutions of
molecules having low molecular masses.

The general model describing the dynamics of systems
characterized by axial degrees of freedom presented here can
be generalized and applied to other systems such as suspensions
of rod-like particles or liquid crystals through an adequate choice of the interaction energy.
This will be done in future work.

\section{Acknowledgements}
We acknowledge useful discussions with Drs.
H. H\'{\i}jar and L. F. del Castillo. JGMB thanks CONACYT for financial support and ISH thanks
UNAM-DGAPA for the financial support of Grant No. IN102609.

\end{document}